\documentclass[11pt]{extarticle}
\usepackage{fullpage}
\usepackage[english]{babel}
\usepackage[utf8]{inputenc}
\usepackage{amsmath, amssymb, bm}
\usepackage{graphicx}
\usepackage{subcaption}
\usepackage[colorinlistoftodos]{todonotes}
\usepackage{indentfirst}

\newcommand{\hiddenpower}[2] { \ifnum \numexpr#2=1 #1 \else #1^#2 \fi }
\numberwithin{equation}{section}

\usepackage{xparse}
\ExplSyntaxOn
\newcounter{diff_order}
\newcounter{diff_power}

\newcommand{\rawdiff}[3]
{
	\setcounter{diff_order}{0}
	\clist_map_inline:nn{#3}{\stepcounter{diff_order}}
	
	\frac{\hiddenpower{#1}{\thediff_order} #2}
	{
		\def\old_var{DefaultValue}
		\setcounter{diff_power}{0}
		
		\clist_map_inline:nn{#3}
		{
			\def\new_var{##1}
			\ifnum \thediff_power=0
				\stepcounter{diff_power}
			\else
				\tl_if_eq:NNTF \new_var \old_var
				{\stepcounter{diff_power}}
				{
					#1 \hiddenpower{\old_var}{\thediff_power}
					\setcounter{diff_power}{1}
				}
			\fi

			\def\old_var{##1}
		}
		
		#1 \hiddenpower{\old_var}{\thediff_power}
	}
}

\ExplSyntaxOff

\newcommand{\lb}{\left(}
\newcommand{\rb}{\right)}

\renewcommand{\cos}[2][1]{\hiddenpower{\text{cos}}{#1} \lb #2 \rb}

\newcommand{\sln}[1]{\ensuremath{\mathfrak{sl}_{#1}}}

\begin{document}


\begin{center}
\strut\hfill



\noindent {\Large{\bf{Non-commutative NLS-type hierarchies: dressing $\&$ solutions}}}\\
\vskip 0.3in

\noindent {\bf {{Anastasia Doikou, Iain Findlay and Spyridoula Sklaveniti}}}
\vskip 0.4in

\noindent {\footnotesize School of Mathematical and Computer Sciences, Department of Mathematics,\\ 
Heriot-Watt University, Edinburgh EH14 4AS, United Kingdom}

\vskip 0.1in
\noindent {\footnotesize {\tt E-mail: a.doikou@hw.ac.uk,\ iaf1@hw.ac.uk,\ ss153@hw.ac.uk }}\\

\vskip 0.60in

\end{center}

\begin{abstract}
\noindent We consider the generalized matrix non-linear Schr\"odinger (NLS) hierarchy. By employing the universal Darboux-dressing 
scheme we derive solutions for the hierarchy of integrable PDEs via solutions of the matrix Gelfand-Levitan-Marchenko equation, 
and we also identify recursion relations that yield the Lax pairs for the whole matrix NLS-type hierarchy. 
These results are obtained considering either matrix-integral or general $n^{th}$ order matrix-differential operators as Darboux-dressing transformations.  
In this framework special links with the Airy and Burgers equations are also discussed.
The matrix version of the Darboux transform is  also examined leading to the non-commutative version of the Riccati equation.
The non-commutative Riccati equation is solved and hence suitable conserved quantities are derived. In this context we also
discuss the infinite dimensional case of the NLS matrix model as it provides a suitable candidate for a quantum version of the usual NLS model. 
Similarly, the non-commutitave Riccati equation for the general dressing transform is derived and it is naturally equivalent to the one emerging 
from the solution of the auxiliary linear problem.

\end{abstract}

\date{}
\vskip 0.4in



\section{Introduction}

\noindent 
Non-linear Schr\"odinger-type  models (AKNS scheme) are among the most well studied prototypical integrable hierarchies (see for instance \cite{AKNS1}--\cite{ADP} and references therein). 
Both continuum and discrete versions have  been widely studied from the point of view of the inverse scattering 
method or the Darboux and Zakharov-Shabat (ZS) dressing methods \cite{ZakharovShabat1}--\cite{Darboux}. Within the ZS scheme \cite{ZakharovShabat2} 
solutions of integrable non-linear PDEs can be obtained by means of solutions of the  associated linear problem, and also the hierarchy of Lax operators can be explicitly constructed. 
Similarly, various studies from the Hamiltonian point of view \cite{FT} in the case of periodic as well as generic integrable boundary conditions (see e.g. \cite{Sklyanin, DoFioRa}) exist.  
In the algebraic scheme the hierarchy of Lax pairs can also be constructed via the universal formula based on the existence of a classical $r$-matrix \cite{STS}, extended also to the case of integrable 
boundary conditions \cite{AvanDoikou_boundary}, as well as at the quantum level \cite{DoikouFindlay}.

We consider in this study a generalized matrix version of an NLS-type model (matrix AKNS version). This investigation can be divided into two main parts: 
\begin{enumerate}
\item We investigate solutions of the generalized 
matrix NLS hierarchy based on the generic notion of Darboux-dressing transformation. We explicitly derive recursion relations that yield the Lax pairs for the whole hierarchy, 
and we also obtain generalized solutions by employing both discrete and continuous solutions of the associated linear problem and solving the matrix (non-commutative) 
Gelfand-Levitan-Marchenko (GLM) equation. Solutions of the generalized mKdV model in particular, a member of the hierarchy under study, are derived and expressed in
terms of Airy functions, generalizing the results in \cite{Ablo2}.
\item We study the non-commutative Riccati flows for the matrix NLS hierarchy, aiming primarily at deriving the corresponding hierarchy of conserved quantities. In particular, 
we derive and solve the non-commutative Riccati equation, and hence we obtain the auxiliary function and relevant  classical and non-commutative conserved quantities. 
In this context the matrix NLS model can be seen as a non-commutative version of the familiar NLS model, and be further upgraded to the quantum version of the usual NLS model provided 
that suitable commutation relations are imposed among the non-commutative fields. It is worth noting that relevant studies on the formulation of the quantum GLM equation 
for the NLS model were discussed in \cite{Thucker}. The dressing transform is also performed in the case of Lax pairs expressed in a matrix form, and the non-commutative 
Riccati  equation for the generic Darboux-dressing is identified.
\end{enumerate}

Let us summarize below what is achieved in this investigation.
In the first part of this study (sections 2 and 3) we implement the generalized Darboux-dressing method in order to derive solutions as well as recursion relations that produce 
the Lax pairs for the whole hierarchy. The Darboux transform in our frame is chosen to be either an integral or a differential operator. When choosing the Darboux transform
to be an integral operator  we basically apply the Zakharov-Shabat (ZS) dressing method as described in \cite{ZakharovShabat2}. One of the main advantages of the ZS dressing method
is that no analyticity conditions are {\it a priori} required for the solutions of the corresponding scattering problem, as is the case when applying the inverse scattering transform,
thus this method provides essentially a linearization of non-linear ODEs and PDEs. Here, based on the universal Darboux-dressing scheme we produce solitonic solutions as well as formal  
expressions for generic solutions via the Fredholm theory (see also \cite{Ablo2}), and also derive sets of recursion relations that yield the Lax pairs for the integrable hierarchy at hand.
Moreover, by exploiting the cubic dispersion relations for the third member of the hierarchy, i.e. the generalized matrix mKdV model, we are able to express the formal solutions in terms of
Airy functions, and show that solutions of the corresponding linear problem satisfy the Airy equation, which is known to be a linearization of Painlev\'e II.
Explicit expressions of the dressed Lax pairs for the first few members of the hierarchy are also reported. The viscous and inviscid matrix 
Burgers equations are also derived as special cases in our setting.

In section 4 we investigate non-commutative Riccati flows associated to the matrix NLS hierarchy; these are also directly related to the notion of 
Grassmannian flows studied in the context of non-local, non-linear PDEs in \cite{BDMS, BDMS2}. We solve the non-commutative Riccati  
equation and subsequently derive suitable conserved quantities based on the solution of both time and space Riccati flows. 
A brief discussion on the  relevance of the quantum version of NLS with our findings is also presented. The Poisson structure  for the 
components of the matrix fields, is identified via the comparison of the  equations of motion from the zero curvature condition and via the Hamiltonian formulation. 
The dressing transform is also performed in the case of Lax pairs expressed in a matrix form, and the equivalence of this description with the case of differential 
operators as dressing transforms is displayed. Moreover, we consider the generic dressing transform expressed as a formal series expansion as well as an
integral operator and identify the associated non-commutative Riccati equation, which is naturally equivalent to the one derived for the solution of the space 
part of the linear auxiliary problem. The findings of this section are closely related to the results presented in \cite{DoikouFindlay} regarding the ideas
on the quantum auxiliary linear problem and the quantum Darboux-B\"acklund transforms. Note that in \cite{DoikouFindlay} a certain version of the 
quantum discrete NLS model was studied, i.e. the quantum Ablowitz-Ladik model, so comparisons with the present findings can be made.

\section{Dressing transformations as integral operators}

\noindent 
The main aims of this section are: 1)  the derivation of solutions of  the integrable non-linear PDEs for the hierarchy under study and 
2) the construction of the Lax pairs of all the members of the matrix NLS (AKNS) hierarchy. The key idea is that 
solutions of the non-linear integrable PDEs are obtained via the matrix Gelfand-Levitan-Marchenko equation, which will be derived below,
 by employing solutions of the associated linear problem.

We review now the main ideas of the generic Darboux-dressing scheme \cite{ZakharovShabat2}. Let
\begin{equation}
{\mathbb D}^{(n)}={\mathbb I}\partial_{t_n} + A^{(n)},~~~~{\mathbb L}^{(n)} = {\mathbb I} \partial_{t_n}+ {\mathbb A}^{(n)},
\end{equation}
where ${\mathbb I}$ in the ${\cal N} \times {\cal N}$ unit matrix and  $A^{(n)},\ {\mathbb A}^{(n)}$ are some 
``bare'' (free of fields) and ``dressed'' operators respectively. In general, they can be ${\cal N} \times {\cal N}$ matrices, 
matrix-differential or matrix-integral operators,
and are related via the generic Darboux-dressing transformation (which is a typical similarity transformation) 
\begin{equation}
{\mathbb G}\ {\mathbb D}^{(n)}= {\mathbb L}^{(n)}\ {\mathbb G} \  \Rightarrow\  
\partial_{t_{n} }{\mathbb G} = {\mathbb G}  A^{(n)} - {\mathbb A}^{(n)} {\mathbb G}. \label{generalDarboux}
\end{equation}
It is worth noting that this fundamental idea was generalized in \cite{BDMS, BDMS2} and led to the construction of non-local  non-linear PDEs as well as to 
their solution via solutions of the associated linear problem.

We consider bare differential operators ${\mathbb D}^{(n)}$ that commute $\big [{\mathbb D}^{(n)},\ {\mathbb D}^{(m)}\big ]=0$, 
then the transformation (\ref{generalDarboux}) leads to the generalized Zakharov-Shabat zero curvature relations, which define the integrable hierarchy
\begin{equation}
\partial_{t_n}{\mathbb A}^{(m)} - \partial_{t_m}{\mathbb A}^{(n)}+\big [{\mathbb A}^{(n)},\ {\mathbb A}^{(m)} \big]=0.
\end{equation}
In our setting here we choose to consider for now $A^{(n)},\  {\mathbb A}^{(n)}$  as matrix-differential operators, 
whereas the dressing transform ${\mathbb G}$ is chosen to be either a  matrix-integral operator or a matrix-differential one.

We first consider the case where the Darboux transform ${\mathbb G}$ is an integral operator, and 
we derive the matrix GLM equation from the fundamental factorization condition (or the general Darboux transform).
Let ${\mathbb Q}_0$ be the solution of the linear problem:
\begin{equation}
{\mathbb Q}_0\ {\mathbb D}^{(n)} = {\mathbb D}^{(n)}\ {\mathbb Q}_0, \label{linearpde}
\end{equation}
and ${\mathbb G}$ is the generic Darboux transform:
\begin{equation}
{\mathbb G}\ {\mathbb Q}_0 = {\mathbb Q}.\label{Darboux0}
\end{equation}
The two equations above lead to ${\mathbb Q}\ {\mathbb D}^{(n)}= {\mathbb L}^{(n)}\ {\mathbb Q}$, which together 
with (\ref{generalDarboux}) naturally suggest the equivalence of the two generic objects ${\mathbb G},\ {\mathbb Q}$.
To ensure invertibility of the related operators we express: ${\mathbb G} = {\mathbb I} + {\mathbb K}^+$, $\ {\mathbb Q}_0 = {\mathbb I} +{\mathbb F}$, 
$\ {\mathbb Q} = {\mathbb I} + {\mathbb K}^{-}$, where ${\mathbb K}^{\pm},\ {\mathbb F}$ have integral representations 
($f$ is the test function, an ${\cal N}$-column vector in general):
\begin{eqnarray}
&& {\mathbb F}(f)(x) =\int_{\mathbb R} F(x,y) f(y)\ dy, \nonumber\\
&& {\mathbb K}^{\pm}(f)(x) =\int_{\mathbb R} K^{\pm} (x,y) f(y)\ dy, \label{kernels}
\end{eqnarray}
such that: $K^{+}(x, y) =0, ~~~x>y,$ and $K^-(x,y) =0, ~~~x<y.$

The operators ${\mathbb K}^{\pm},\ {\mathbb F}$ then satisfy via (\ref{Darboux0}) the factorization condition 
\begin{equation}
{\mathbb K}^+  +{\mathbb F}+ {\mathbb K}^+ {\mathbb F} = {\mathbb K}^-, \label{factorization}
\end{equation}
which leads to the fact that the kernel $K^+(x,y)$ satisfies the Gelfand-Levitan-Marchenko equation 
and $K^-(x, y)$ obeys an analogous integral equation:
\begin{eqnarray}
&& K^{+}(x,z) +F(x,z) + \int_x^{\infty}dy\ K^+(x,y)F(y,z)=0, ~~~~z>x, \nonumber\\
&& K^-(x, z) = F(x,z) + \int_x^{\infty}dy\ K^+(x,y)F(y,z), ~~~~~z<x.
\end{eqnarray}
$F(x, y)$ is the solution of the linear problem, i.e. invariance of the differential operators ${\mathbb D}^{(n)}$ under the action of the operator ${\mathbb F}$ is required (\ref{linearpde}).
Dependence on the universal time $t$ that includes all the times $t_n$ is implied but omitted for now for simplicity.
Note that the factorization condition (\ref{factorization}) is nothing but the analogue of the Darboux-dressing transformation acting on the linear solution  ${\mathbb F}$ and providing 
the transformed solution ${\mathbb K}^-$.  We shall address this issue again when dealing with the auxiliary matrix problem in the matrix language. 
The kernel $K^+$, as will be transparent in the following,  is the quantity that produces solutions of the integrable PDEs emerging from the zero curvature condition. 
One may think of the GLM equation as a necessary intermediate  step between the linear problem and  the non-linear integrable PDE.

It is worth pointing out that the integral representations (\ref{kernels}) lead to a Volterra type integral equation (GLM). 
In this frame the condition (\ref{Darboux0}) is equivalent to a Borel decomposition, but we rather prefer to consider analytical arguments based on the 
type of kernels of integral operators in this section generalizing the description in \cite{ZakharovShabat1, ZakharovShabat2, Ablo2}. The Volterra type integral equations 
guarantee the existence of  boundary terms that provide the relevant local fields, as will be transparent in subsection 2.2. 
In general, one may consider instead of Volterra integral equations, 
Fredholm type integral equations as in \cite{BDMS, BDMS2}, and obtain non-local, non-linear integrable PDEs.
The dressing relations (\ref{generalDarboux}) uniquely fix the Darboux integral transform as well as the solutions, provided that specific solutions
of the associated linear problem are available, as will be discussed in subsections 2.1 and 2.2. Lets us also note that the GLM method is equivalent to the Riemann problem based on the typical factorization problem of matrix valued functions. 
For a detailed, although rather technical discussion on this correspondence we refer the interested reader to \cite{FT}.

We focus now on the sets of operators associated to the generalized matrix NLS model:
\begin{eqnarray}
&&{\mathbb D}^{(0)} = {\cal W} \partial_x, ~~~~{\mathbb D}^{(1)} = {\mathbb I} \partial_{t_1}- \hat {\cal W}  \partial_x, 
~~~~{\mathbb D}^{(n)} ={\mathbb I} (\partial_{t_n} - \partial_x^n) , ~~~n>1, \label{DD}\\
&& {\mathbb L}^{(0)} = {\cal W} \partial_x + U(x), ~~~~~{\mathbb L}^{(1)} = {\mathbb I} \partial_{t_1}- \hat {\cal W} \partial_x + a(x), 
~~~~{\mathbb L}^{(n)} ={\mathbb I} (\partial_{t_n} - \partial_x^n )+
 \sum_{k=0}^{n-1} a_k(x) \partial_x^k.  \nonumber\\ &&\label{LL}
\end{eqnarray}
$U =\begin{pmatrix}
 0& \hat u \cr
u &0
\end{pmatrix}$,  ${\mathbb I}$ is the ${\cal N} \times {\cal N}$ identity matrix ($N+M={\cal N}$),
\begin{eqnarray}
&& {\cal W} = \begin{pmatrix}
  w_1\ {\mathbb I}_{N\times N}& 0_{N\times M}\cr
 0_{M\times N} & w_2\ {\mathbb I}_{M\times M}
\end{pmatrix}.
\end{eqnarray}
Similarly $\hat {\cal W}$ is defined as above with constants $\hat w_1,\  \hat w_2$ ($w_1 \neq w_2,\ \hat w_1\neq \hat w_2$ are arbitrary constants).
The quantities $U(x),\ a(x),\ a_k(x)$ are going to be identified via the dressing process. In general, ${\cal W}$ can be any ${\cal N}\times{\cal N}$ matrix, 
but we focus henceforth on the case where it is given in the block form above. 
Different choices of the ${\cal W}$ matrix will naturally give rise to models with distinct underlying symmetries.

Note that the linear PDEs  (\ref{linearpde}) lead to linear differential equations for all the involved time flows
\begin{eqnarray}
&& {\cal W}\partial_x F(x, y) + \partial_y F(x,y){\cal W}=0, \nonumber\\
&& \partial_{t_n} F(x,y)-\partial^n_x F(x,y) + (-1)^n \partial_y^n F(x,y)=0, \label{linear}
\end{eqnarray}
the dependence of the functions on time $t_n$ is omitted for brevity.  Let us now express the solutions of the linear problem for the matrix NLS as
\begin{eqnarray}
&& F(x,z) =  \begin{pmatrix}
 0_{N\times N} & f_{N\times M}(x,y)\cr
 \hat f_{M\times N}(x,y) & 0_{M\times M}
\end{pmatrix}. \label{GrasF} 
\end{eqnarray}
This will be used subsequently for obtaining solutions of the GLM equation and hence solutions for the tower of integrable non-linear
PDEs of the matrix NLS hierarchy\footnote{Note that in the special case where $w_1=-w_2$,  $N= M$ and $u=\hat u$, $f = \hat f$ 
one recovers the matrix mKdV model. In the familiar (imaginary time) matrix NLS model $\hat u = u^{\dag}$ and $\hat f = f^{\dag}$, while in the case 
where $u=1$ one recovers the matrix KdV hierarchy. Generally speaking we are dealing here with the Grassmannian $Gr(N|{\cal N})$ AKNS scheme, 
but we are primarily interested in the case where $u\neq \hat u$, and also when both $u, \hat u$ are non-constant matrices, hence the appellation matrix NLS-type model.}.

\subsection{Solutions of the matrix GLM equation}

\noindent The first step is the derivation of solutions of the GLM equation by means of the linear solutions above (\ref{GrasF}). Let the
matrix kernel $K^+$ be expressed as 
\begin{equation}
 K^+(x,y )=\begin{pmatrix}
{\mathbb A}_{N\times N}(x,y)& {\mathbb B}_{N\times M}(x,y)\cr
{\mathbb C}_{M\times N}(x,y)& {\mathbb D}_{M\times M}(x,y)
\end{pmatrix}. \label{GrasK0}
\end{equation}
Inserting the matrix expressions  (\ref{GrasF}) and (\ref{GrasK0}) into the GLM equation we obtain two independent sets of equations involving 
the matrix fields ${\mathbb A},\ {\mathbb B}$
\begin{eqnarray}
&& {\mathbb B}(x,z) + f(x,z) + \int_x^{\infty}dy\ {\mathbb A}(x,y)  f(y,z)=0, \label{I1}\\
&& {\mathbb A}(x,z) + \int_x^{\infty}dy\ {\mathbb B}(x,y) \hat f(y,z)=0, \label{I}
\end{eqnarray}
and the fields ${\mathbb C},\ {\mathbb D}$
\begin{eqnarray}
&& {\mathbb C}(x,z) +\hat f(x,z) + \int_x^{\infty}dy\ {\mathbb D}(x,y)\hat   f(y,z)=0, \label{II1}\\
&& {\mathbb D}(x,z) + \int_x^{\infty}dy\ {\mathbb C}(x,y)  f(y,z)=0.\label{II}
\end{eqnarray}
These two sets are independently solved and the matrix-fields ${\mathbb B}$ and ${\mathbb C}$ are then given by the expressions:
\begin{eqnarray}
&& {\mathbb B}(x,z) +f(x,z)- \int_x^{\infty}d\tilde y\ \int_x^{\infty}dy\  {\mathbb B}(x, \tilde y)\hat f(\tilde y, y)f(y,z) =0, \nonumber\\
&& {\mathbb C}(x,z) + \hat f(x,z)- \int_x^{\infty}d\tilde y\ \int_x^{\infty}dy\  {\mathbb C}(x, \tilde y)f(\tilde y, y)\hat f(y, z)=0.  \label{fields}
\end{eqnarray}
The latter equations can be expressed in a generic operatorial form as
\begin{eqnarray}
&& {\mathfrak B} = - {\mathfrak f}\ {\mathfrak g}^{-1}, ~~~~~{\mathfrak C} = - \hat {\mathfrak f}\ \hat  {\mathfrak g}^{-1} \label{abstract}
\end{eqnarray}
provided that $ {\mathfrak g},\  \hat  {\mathfrak g}$ are invertible, and we define
\begin{equation}
{\mathfrak g} ={\tt id} - \hat {\mathfrak f} {\mathfrak f}, ~~~~~\hat {\mathfrak g} ={\tt id} - {\mathfrak f} \hat {\mathfrak f}. \label{solutionsa}
\end{equation}
${\mathfrak B},\ {\mathfrak C}$ are integral operators with kernels
${\mathbb B}(x, y),\ {\mathbb C}(x, y)$, and ${\mathfrak g},\ \hat {\mathfrak g}$ are also integral operators with kernels 
\begin{equation}
g(x; y, z) =\delta(y-z) - \int_{x}^{\infty}d\tilde y\  \hat f(y, \tilde y) f(\tilde y, z), ~~~~\hat g(x; y, z) =\delta(y-z)- \int_{x}^{\infty}d\tilde y\  f(y, \tilde y) \hat f(\tilde y, z)
\end{equation}
respectively, as dictated by (\ref{fields}).  Note also that the inverse of any operator of the form $\mbox{id} -{\mathfrak h}$ can be formally expressed as 
\begin{equation}
(\mbox{id} -{\mathfrak h})^{-1} = \mbox{id} + \sum_{k=1}^{\infty}{\mathfrak h}^{k}. \label{inverse}
\end{equation}

The solutions ${\mathbb B}, {\mathbb C}$, expressed exclusively in terms of solutions of the linear problem $f,\ \hat f$ (\ref{linear}), produce the fields of the matrix NLS hierarchy as will be shown in a subsequent 
section when constructing the associated Lax pairs via the dressing transform. We introduce below both discrete and continuum solutions of the linear equations
and obtain the corresponding  solutions of the matrix GLM equation.

\subsubsection*{A. Discrete solutions}

\noindent  
We first consider discrete solutions of the linear problem, which can be expressed as
\begin{equation}
f(x,z, t) = \sum_{\alpha=1}^L b_{\alpha} e^{\sum_n \Lambda_{\alpha}^{(n)}t_n - \kappa_{\alpha}x - \mu_{\alpha}z},
~~~~~\hat f(x,z,t) = \sum_{\alpha=1}^L \hat b_{\alpha} e^{\sum_n \hat \Lambda_{\alpha}^{(n)}t_n - \hat \mu_{\alpha}x - \hat \kappa_{\alpha}z}.
\end{equation}
We make the dependence on $t = \{t_n\}$ (the ``universal'' time containing all the times $t_n$) explicit  in order to show the dispersion relations for each time flow.
Indeed, the general dispersion relations immediately follow from (\ref{linear}), and are given as
\begin{eqnarray}
&& w_1\kappa_{\alpha}+ w_2 \mu_{\alpha} =0, ~~~~w_1 \hat \kappa_{\alpha} + w_2 \hat \mu_{\alpha} =0, \nonumber\\
&&  \Lambda_{\alpha}^{(n)} -(-1)^n \kappa_{\alpha}^n + \mu_{\alpha}^n=0, ~~~
~\hat \Lambda_{\alpha}^{(n)} -(-1)^n \hat \mu^n_{\alpha} + \hat \kappa_{\alpha}^n =0. \label{disp1}
\end{eqnarray}

Taking into consideration the form of the solutions of the linear problem as well as equation (\ref{fields}) we consider the generic form for the matrix fields:
\begin{eqnarray}
{\mathbb B}(x,z,t) = \sum_{\alpha=1}^L {\mathbb L}_{\alpha}(x,t)e^{-\mu_{\alpha} z},
~~~~~{\mathbb C}(x,z,t) = \sum_{\alpha=1}^L \hat{ \mathbb L}_{\alpha}(x,t)e^{-\hat \kappa_{\alpha} z}.
\end{eqnarray}
Let us also introduce some important objects:  ${\mathbb M}$ and $\hat {\mathbb M}$ are operator valued matrices with elements  ${\mathbb M}_{\alpha \beta},\  
\hat {\mathbb M}_{\alpha \beta}$  being themselves $M\times M$ and $N \times N$ matrices respectively  defined as:
\begin{eqnarray}
&&{\mathbb M}(x,t) = {\mathbb I}_{M\times M} \otimes {\mathbb I}_{L\times L} -{\mathbb P}(x,t), ~~~~~~
\hat {\mathbb M}(x,t) = {\mathbb I}_{N\times N} \otimes {\mathbb I}_{L\times L} -\hat {\mathbb P}(x,t), \nonumber\\
&& {\mathbb P}_{\beta \alpha }(x,t) =\sum_{\gamma=1}^L {\mathrm f}_{\beta \gamma \alpha}(x,t)\hat b_{\gamma} b_{\alpha}, 
~~~~\hat {\mathbb P}_{ \beta \alpha}(x,t) = \sum_{\gamma=1}^L \hat {\mathrm f}_{\beta \gamma \alpha}(x,t) b_{\gamma}\hat  b_{\alpha},
\end{eqnarray}
and we also define
\begin{eqnarray}
&& {\mathrm f}_{\beta\gamma\alpha}(x,t)= e^{\sum_n(\hat \Lambda_{\gamma}^{(n)}+ \Lambda_{\alpha}^{(n)})t_n}
{e^{-(\mu_{\beta} + \hat \mu_{\gamma})x} e^{-(\hat \kappa_{\gamma} + \kappa_{\alpha} )x}\over
 (\mu_{\beta} + \hat \mu_{\gamma})(\hat \kappa_{\gamma} + \kappa_{\alpha} ) }, \nonumber\\
&& \hat {\mathrm f}_{\beta\gamma\alpha}(x,t)= e^{\sum_n( \Lambda_{\gamma}^{(n)}+\hat \Lambda_{\alpha}^{(n)})t_n}
{e^{-(\hat \kappa_{\beta} +  \kappa_{\gamma})x} e^{-(\mu_{\gamma} + \hat \mu_{\alpha} )x}\over 
(\hat \kappa_{\beta} + \kappa_{\gamma} ) (\mu_{\gamma} +\hat \mu_{\alpha}) }.
\end{eqnarray}
Then the matrices ${\mathbb L},\ \hat {\mathbb L}$ are identified as
\begin{eqnarray}
{\mathbb L}(x, t){\mathbb M}(x, t)= -{\mathrm B}(x, t) , ~~~~~\hat {\mathbb L}(x, t) \hat {\mathbb M}(x, t) = -\hat {\mathrm B}(x,t). \label{linear-matrix}
\end{eqnarray}
${\mathrm B},\ \hat{\mathrm  B}$ are operator valued $L$-vectors with components: $b_{\alpha} e^{\sum_n\Lambda^{(n)}_{\alpha} t_n- \kappa_{\alpha}x} $, 
$\hat b_{\alpha} e^{\sum_n\hat \Lambda^{(n)}_{\alpha} t_n- \hat \mu_{\alpha}x }$ respectively. Provided that ${\mathbb M},\  \hat {\mathbb M}$ are invertible, 
i.e $\det({\mathbb I} - {\mathbb P})\neq 0,\ \det({\mathbb I} - \hat {\mathbb P})\neq 0$, 
then the general discrete solutions can be expressed in a formal series expansion as
\begin{eqnarray}
&& {\mathbb L}(x,t) = - {\mathrm B}(x,t)\Big ({\mathbb I}_{M\times M} \otimes {\mathbb I}_{L\times L} + \sum_{m=1}^{\infty} {\mathbb P}^m\Big ). \label{solution-matrix}
\end{eqnarray}
Similarly, for $\hat {\mathbb L}$: ${\mathrm  B} \to \hat {\mathrm  B} $, ${\mathbb P} \to \hat {\mathbb P}$ and ${\mathbb I}_{M \times M} \to {\mathbb I}_{N\times N}$.

It will be instructive to provide the explicit expression for the ``one-soliton'' solution $L=1$. Equations (\ref{linear-matrix})  in this case reduce to
\begin{eqnarray}
&& {\mathbb L}(x,t) \Big ({\mathbb I}_{M\times M} - {\mathrm f}(x,t) \hat b b \Big) = - b e^{\sum_n\Lambda^{(n)}t_n - \kappa x }, \nonumber\\
&& \hat {\mathbb L}(x,t) \Big ({\mathbb I}_{N\times N} - {\mathrm f}(x,t) b\hat b \Big) = -\hat b  e^{\sum_{n} \hat \Lambda^{(n)}t_n - \hat \mu x }, \nonumber\\
&& {\mathrm f}(x,t) = e^{\sum _n(\Lambda^{(n)} + \hat \Lambda^{(n)}) t_n }{e^{-(\mu+ \hat \mu)x} e^{-(\hat \kappa+ \kappa )x}\over (\mu+ \hat \mu)(\hat \kappa + \kappa ) },
\end{eqnarray} 
then the fields ${\mathbb B},\  {\mathbb C}$ are given in a compact form as
\begin{eqnarray}
&& {\mathbb B}(x,z, t) = -b\Big ({\mathbb I}_{M \times M}+  g(x,t) \hat b b\Big ) e^{\sum_n \Lambda^{(n)}t_n - \kappa x - \mu z}, \\
&& {\mathbb C}(x,z,t) = -\hat b \Big ({\mathbb I}_{N \times N}+  \hat g(x,t) b \hat  b\Big ) e^{\sum_n\hat  \Lambda^{(n)}t_n - \hat \mu x - \hat \kappa z},
\end{eqnarray}
where $g= {{\mathrm f}\over 1-\xi {\mathrm f} }$ and we have assumed $(b\hat b)^2 = \xi b\hat b$, (see also relevant results for the vector NLS in \cite{Ablo, ADP}).
The kernels  ${\mathbb A},\  {\mathbb D}$ of the matrix kernel $K^+$ (\ref{GrasK}) can be identified by means of relations (\ref{I1}) and (\ref{II}) respectively.
If we further impose Temperley-Lieb type constraints for the quantities $b,\ \hat b$: $\ b \hat b b=\xi b\ $ and $\ \hat b b \hat b=\xi \hat b$,
we conclude,
\begin{eqnarray}
&& {\mathbb B}(x,z,t) = -b { e^{\sum_n \Lambda^{(n)}t_n - \kappa x - \mu z} \over 1-\xi {\mathrm f}}, \\
&& {\mathbb C}(x,z,t) = -\hat b{  e^{\sum_n\hat  \Lambda^{(n)}t_n - \hat \mu x - \hat \kappa z}\over 1-\xi  {\mathrm f}}.
\end{eqnarray}
As will become transparent later in the text  ${\mathbb B}(x,x),\  {\mathbb C}(x,x)$ are proportional to the relevant NLS matrix fields. 
It is worth mentioning that for the one soliton case the constraint is actually
justified as a consistency condition when computing also the quantities ${\mathbb A},\ {\mathbb D}$ via equations (\ref{I}) and (\ref{II}).

\subsubsection*{B. Continuous Solutions}

\noindent We come now to the more general scenario of continuum solutions of the linear problem. More precisely, the generic solution of the linear problem can be expressed 
as a continuum Fourier transform
\begin{eqnarray}
&& f(x,z, t) = \int_{\mathbb R} dk\ b_k \exp\Big [\sum_n \Lambda_k^{(n)}t_n  +ik x + i\mu_k z\Big ],\nonumber\\
&& \hat f(x,z, t) = \int_{\mathbb R} d k\  \hat  b_{ k} \exp\Big [\sum_n \hat \Lambda_{ k}^{(n)}t_n  +i\mu_k x + ik z\Big ],
\end{eqnarray}
where the coefficients $b$ and $\hat b$ are $N\times M$ and $M\times N$ matrices respectively.
As in the case of discrete solutions the dispersion relations then immediately follow from (\ref{linear}):
\begin{eqnarray}
w_1k+ w_2 \mu_k=0, ~~~ \Lambda_k^{(n)} -i^n k^n +(-i)^n \mu_k^n=0, ~~~~\hat \Lambda_k^{(n)} -i^n \mu_k^n+(-i)^n k^n=0. \label{disp2}
\end{eqnarray}
Taking into account the solutions of the linear problem as well as equation (\ref{fields}) we consider for the matrix fields:
\begin{eqnarray}
{\mathbb B}(x,z,t) =\int_{\mathbb R} dk\ {\mathbb L}(k, x,t)e^{i\mu_{k} z},~~~~~{\mathbb C}(x,z, t) = \int_{\mathbb R} dk\  \hat{ \mathbb L}(k,x,t_n)e^{i \hat kz}.
\end{eqnarray}
Let us also define the quantities:
\begin{eqnarray}
&& {\mathrm B}(k,x,t)= b_{k} e^{\sum_n\Lambda^{(n)}_{k} t_n+ i  k x }, ~~~~\hat {\mathrm B}(k,x, t) = \hat b_{k} e^{\sum_n\hat \Lambda^{(n)}_{k} t_n+i\mu_k x}, \nonumber\\
&& {\mathbb P}(k_1, k, x,t) = \int_{\mathbb R} dk_2\ {\mathrm f}(k_1,k_2,k, x,t)\hat b_{k_2} b_{k}, \nonumber\\
&& \hat {\mathbb P}(k_1, k, x,t) = \int_{\mathbb R} dk_2\ \hat  {\mathrm f}(k_1,k_2,k, x,t)b_{k_2} \hat b_{k}
\end{eqnarray}
recall $\mu_k = - {w_1 \over w_2}k$, and 
\begin{eqnarray}
&& {\mathrm f}(k_1,k_2,k,x,t)= - e^{\sum_n(\hat \Lambda_{k_2}^{(n)}+ \Lambda_{k}^{(n)})t_n}{e^{i(\mu_{k_1} + \mu_{k_2})x} e^{i(k_2+ k )x}\over 
(\mu_{k_1} + \mu_{k_2})(k_2+ k ) }, \nonumber\\
&& \hat {\mathrm f}(k_1,k_2,k,x,t)= - e^{\sum_n(\Lambda_{k_2}^{(n)}+ \hat \Lambda_{k}^{(n)})t_n}{e^{i(k_1+k_2)x} e^{i(\mu_{k_2} +
\mu_{k} )x}\over (k_1+k_2)(\mu_{k_2} + \mu_k ) }.
\end{eqnarray}
Provided that  the operators ${\mathbb I} - {\mathbb P},\  {\mathbb I} - \hat {\mathbb P}$ are invertible  i.e. the Fredholm determinants are non-zero
($\det({\mathbb I} - {\mathbb P})\neq 0,\ \det({\mathbb I} - \hat {\mathbb P})  \neq 0$),
then the matrices ${\mathbb L},\ \hat {\mathbb L}$ are explicitly identified  via the  integral equations
\begin{eqnarray}
&& \int dk_1 {\mathbb L}(k_1, x,t)\Big  ({\mathbb I}_{M\times M}\delta(k_1, k) - {\mathbb P}(k_1, k, x,t)\Big)= -{\mathrm B}(k,x, t),
\end{eqnarray}
similarly, for $\hat {\mathbb L}$: ${\mathrm  B} \to \hat {\mathrm  B} $, ${\mathbb P} \to \hat {\mathbb P}$ and ${\mathbb I}_{M \times M} \to {\mathbb I}_{N\times N}$.

The latter relations provide the formal series expansion for ${\mathbb L}$ and $\hat{\mathbb L}$, i.e. 
the integral analogues of the matrix relations (\ref{solution-matrix}) presented in the discrete case previously:
\begin{eqnarray}
&& {\mathbb L}(k) =-{\mathrm  B}(k) - \sum_{m=1}^{\infty}\int_{\mathbb R} dk_1 \ldots \int_{\mathbb R} dk_m\  
{\mathrm B}(k_1) {\mathbb P}(k_1, k_2){\mathbb P}(k_2, k_3) \ldots {\mathbb P}(k_m,k), \nonumber\\
&&
\end{eqnarray}
similarly for $\hat {\mathbb L}$: ${\mathrm  B} \to \hat {\mathrm  B} $ and ${\mathbb P} \to \hat {\mathbb P}$; dependence of $x,\  t$ is implied in the expression above.
The latter is the analogue of (\ref{solutionsa}), (\ref{inverse}) expressed in terms of the kernels of the integral operators.

\begin{itemize}

\item {\tt Generalized mKdV solutions $\&$ Airy functions}\\
It is interesting to focus now on solutions of the generalized mKdV equation ($n=3$). The key observation is that due to the cubic dispersion relations in the case $n=3$ (\ref{disp2}) 
the solutions of the linear problem can be expressed in terms of Airy functions $\mbox{Ai}(x)$. Indeed, after expressing the coefficients $b_k,\ \hat b_k$ in terms of the initial values 
of the solutions of the linear problem $f_0,\ \hat f_0$ at $t=0$, via an inverse Fourier transform
\begin{equation}
b_k = {1\over 2\pi}\int_{\mathbb R} d\xi\  f_0(\xi)e^{-ik\xi}, ~~~~\hat b_k = {1\over 2\pi}\int_{\mathbb R} d\xi\  \hat f_0(\xi)e^{-ik\xi},
\end{equation}
and recalling the definition of the Airy function
\begin{equation}
\mbox{Ai}(x) = {1\over \pi}\int_{0}^{\infty}dt\ \cos{{t^3\over 2} +xt},
\end{equation}
we obtain
\begin{equation}
f(x,z) ={1\over \nu} \int_{\mathbb R} d\xi\ \mbox{Ai}({x+sz-\xi \over \nu})f_0(\xi), ~~~~\hat f(x,z) ={1\over \nu} \int_{\mathbb R} d\xi\ \mbox{Ai}({sx+z-\xi \over \nu})\hat f_0(\xi),
\end{equation}
where $s= -{w_1\over w_2},\ \nu = (3(1-s^3)t)^{1\over3}$. 

We can express the  kernels appearing in the 
fundamental expressions (\ref{fields})  in terms of Airy functions, indeed define
$\hat {\cal F}(x; \tilde y, z) = \int_{x}^{\infty} dy\ f(\tilde y,y)\hat f(y,z)$ and  
${\cal F}(x; \tilde y, z) = \int_{x}^{\infty} dy\ \hat f(\tilde y,y)f(y,z)$, then
\begin{eqnarray}
&&\hat  {\cal F}(x; \tilde y,z) = {1\over \nu^2} \int_{x}^{\infty} dy\int_{\mathbb R} d\xi \int_{\mathbb R} d \tilde \xi\ \mbox{Ai}({\tilde y+sy-\xi \over \nu}) 
\mbox{Ai}({sy+z-\tilde \xi \over \nu} )f_0(\xi)  \hat f_0(\tilde \xi), \nonumber\\
&& {\cal F}(x;\tilde y,z) = {1\over \nu^2} \int_{x}^{\infty} dy \int_{\mathbb R} d\xi \int_{\mathbb R} d \tilde \xi\ \mbox{Ai}({s\tilde y+y-\xi \over \nu}) 
\mbox{Ai}({y+sz-\tilde \xi \over \nu}) \hat f_0(\xi)  f_0(\tilde \xi). \nonumber\\
&& \label{Airy1}
\end{eqnarray}
Distinct  choices of the initial functions $f_0,\ \hat f_0$  give rise to different generic solutions, but an obvious choice of initial conditions is for instance, 
$f_0(\xi) =\delta(\xi)\ {\mathrm m}_{N\times M}$, $\  \hat f_0(\xi) = \delta(\xi)\ \hat {\mathrm m}_{M\times N}$. 
Given that $f,\ \hat f$ satisfy the linear problem (\ref{linear}) for $n=3$, it is straightforward to see that both $\nu f({x+sz\over \nu}),\ \nu \hat f({sx+z\over \nu}) $ 
satisfy as expected the Airy equation, (linearization of Painlev\'e II (see also \cite{Ablo2}))
\begin{equation}
{\partial^2 F(\zeta) \over \partial \zeta^2}- \zeta F(\zeta) =0,
\end{equation}
with the parameter $\zeta$ defined as ${x+sz\over \nu}$ and ${sx +z \over \nu}$ respectively, and having also assumed that $F(\zeta) \to 0$ when $\zeta \to \infty$. 
The matrix fields ${\mathbb B},\  {\mathbb C}$ can be then expressed as a formal series expansion in terms of Airy functions via (\ref{fields}) and (\ref{Airy1}).
These findings  are in tune with the results derived  in  \cite{Ablo2}, where the connection between the mKdV and Painlev\'e II are discussed. 
Similar observations can be made for the matrix NLS model ($n=2$) provided that the values of $w_1,\ w_2$ are suitably tuned. Indeed, in this case the solutions 
can be expressed in terms of the heat kernel. We shall comment in the next subsection on the matrix Burgers equation as a special case of our construction. 
As is well known in the viscous Burgers case, solutions can be obtained in terms of the heat kernel via the Cole-Hopf transformation.
\end{itemize}

\subsection{Dressing of linear operators: the hierarchy}

\noindent 
The next important task is to obtain the hierarchy of the Lax  pairs via the dressing process (\ref{generalDarboux}), 
where ${\mathbb G}= {\mathbb I} +{\mathbb K}^+$ is chosen to be an integral operator with kernel given in (\ref{kernels}). 
Note that in the rest of this section dependence on the universal time $t$ is implied, but omitted for simplicity. 
Let us first apply the dressing transformation for the ${\mathbb L}^{(0)}$ 
(the $t$ independent part of the Lax pair):
\begin{equation}
{\mathbb G}\ {\cal W} \partial_x = \big ({\cal W}\partial_x + U\big )\ {\mathbb G},
\end{equation}
which directly leads to 
\begin{eqnarray}
\int_{x}^{\infty}dy\ \Big ( {\cal W}\partial_x K(x,y) &+& \partial_y K(x,y) {\cal W} + U(x) K(x,y)\Big )f(y) \nonumber\\
&+& \Big( U(x)+ K(x,x){\cal W}
-{\cal W}K(x,x)\Big)f(x) =0,
\end{eqnarray}
yielding the fundamental equations
\begin{eqnarray}
&& U(x) = {\cal W}K(x,x) - K(x,x) {\cal W}, \nonumber\\
&&  {\cal W}\partial_x K(x,y) + \partial_y K(x,y)  {\cal W} + U(x) K(x,y)=0.
\end{eqnarray}
The latter equations provide the form of $U$  as well the $K$-matrix
\begin{equation}
U(x)=  \begin{pmatrix}
		0 & \hat u(x) \\
	u(x)   & 0
	\end{pmatrix}, \label{UU}
\end{equation}
\begin{equation}
\hat u(x) = h{\mathbb B}(x,x), ~~~~u(x) = -h{\mathbb C}(x,x), ~~~\mbox{and} ~~~h = w_1 -w_2, \label{fields1}
\end{equation}
and the set of constraints 
\begin{eqnarray}
&& w_1 \Big (\partial_x {\mathbb A}(x,y) + \partial_y {\mathbb A}(x,y)\Big ) = -h{\mathbb B}(x,x){\mathbb C}(x,y), \nonumber\\
&&  w_2 \Big (\partial_x {\mathbb D}(x,y) + \partial_y {\mathbb D}(x,y)\Big ) = h{\mathbb C}(x,x){\mathbb B}(x,y), \nonumber\\
&& w_1 \partial_x {\mathbb B}(x,y) +w_2 \partial_y {\mathbb B}(x,y)= -h{\mathbb B}(x,x){\mathbb D}(x,y),  \nonumber\\
&& w_2 \partial_x {\mathbb C}(x,y) +  w_1\partial_y {\mathbb C}(x,y) = h{\mathbb C}(x,x){\mathbb A}(x,y). \label{constr1}
\end{eqnarray}

Let us now implement the dressing transform for all the associated time flows $t_n,$ and derive the dressed operators ${\mathbb L}^{(n)},\  n>0$, (\ref{LL}).
\begin{itemize}

\item ${\bf n=1}$: equation (\ref{generalDarboux}) yields
\begin{eqnarray}
\int_x^{\infty}dy\ K(x,y)\big (\partial_{t_1}-\hat {\cal W} \partial_y \big )f(y) &=& a(x) f(x) +\big  (\partial_{t_1}-\hat {\cal W} \partial_x\big )\int_x^{\infty}dy\ K(x,y)f(y) 
\nonumber\\ &+&\int_x^{\infty}dy\ a(x) K(x,y)f(y).
\end{eqnarray}
After integrating by parts and considering the boundary terms we conclude:
\begin{eqnarray}
&& a(x) =  K(x,x)\hat {\cal W} - \hat {\cal W}K(x,x),  \nonumber\\
&&  \partial_{t_1}K(x,y)= \hat {\cal W}\partial_x K(x,y) + \partial_y K(x,y)  \hat {\cal W} - a(x) K(x,y).
\end{eqnarray}

\item ${\bf  n>1}$: from the basic dressing relation (\ref{generalDarboux}) we obtain 
\begin{eqnarray}
\int_x^{\infty}dy\ K(x,y)\big (\partial_{t_n}-\partial_y^n\big )f(y) &=& {\mathbb X}_x f(x) +\big  (\partial_{t_n}-\partial_x^n\big )\int_x^{\infty}dy\ 
K(x,y)f(y) \nonumber\\ &+&\int_x^{\infty}dy\ {\mathbb X}_xK(x,y)f(y), 
\end{eqnarray}
where we define ${\mathbb X}_x = \sum_{k=0}^{n-1}a_k(x) \partial_x^k$. After repeated integrations by parts, and carefully taking into consideration 
the boundary terms by iteration, we conclude:
\begin{eqnarray}
&& \int_x^{\infty}dy\ \Big ( \partial_{t_n} K(x,y)  - \partial_x^n K(x,y )+ (-1)^n \partial_y^n K(x,y )+{\mathbb X}_x K(x,y)\Big ) f(y) \nonumber\\
&&+\sum_{k=0}^{n-1} a_k \partial_x^k f(x) - \sum_{m=0}^{n-1}(-1)^m \partial_y^m K(x,y)|_{x=y}\ \partial_x^{n-m-1}f(x) + \partial_x^{n-1}\big  (K(x,x) f(x)\big ) \nonumber\\
&& + \sum_{m=0}^{n-2} \partial_x^m\big (\partial_x^{n-m-1}K(x,y)|_{x=y}f(x)\big ) - \sum_{k=1}^{n-1} a_k(x) \partial_x^{k-1} \big (K(x,x)f(x)\big ) \nonumber\\ 
&&- \sum_{k=1}^{n-1} a_k(x)\sum_{m=0}^{k-2}
 \partial^m_x\big (\partial_x^{k-m-1}K(x,y)|_{x=y} f(x)\big ) =0. \label{dress1}
\end{eqnarray}
It follows from the expression above that $K$ satisfies:
\begin{equation}
 \partial_{t_n} K(x,y)  - \partial_x^n K(x,y )+ (-1)^n \partial_y^n K(x,y )+\sum_{k=0}^{n-1} a_k(x) \partial^k_x K(x,y) =0, \label{dress2}
\end{equation}
whereas the use of the boundary terms for each order $\partial_x^m f(x)$ provides the elements $a_k$. In fact, similar relations arise when considering a 
differential operator as the Darboux dressing transform, as is discussed in the next subsection. 

\item {\tt Matrix Burgers equation} \\
Let us now focus on the linear operator ${\mathbb I}(\partial_t -\partial^2_x)$ independently of the existence of a Lax pair.
The dressed operator is required to be of the form ${\mathbb I} (\partial_t -\partial^2_x) +V(x)$, then 
via the dressing process described by (\ref{dress1}), (\ref{dress2}), we conclude ($n=2$):
\begin{eqnarray}
&& \partial_t K(x, y) - \partial_x^2 K(x, y) +  \partial_y ^2 K(x, y) + V(x) K(x, y) =0, \nonumber\\ 
&&  V(x) = -2\Big  (\partial_x K(x,y) + \partial_y K(x,y)\Big )\Big |_{x=y},
\end{eqnarray}
where $V$ and $K$ are generic ${\cal N} \times {\cal N}$ matrices.
Let us also consider the following general condition for the kernel: $K(x, y) = K(x+wy)$, $w$ is a constant, 
then the equations above produce at $x=y$ and after setting $\chi = (1+w) x$:
\begin{equation}
\partial_t K(\chi) +(w^2 -1)\partial^2_{\chi}K(\chi) -2 (1+w) \partial_{\chi}K(\chi)\ K(\chi) =0.
\end{equation}

The latter yields  both the viscous and inviscid matrix Burgers equations. Indeed, for $w=1$, and
after rescaling the time $\tau = -4 t$ we obtain the inviscid Burgers equation
\begin{equation}
\partial_{\tau} K(\chi) + \partial_{\chi}K(\chi) K(\chi) =0.
\end{equation}
For $w\neq \pm 1$, and after setting $\tau = -2(1+w) t $ and $\nu ={w-1 \over 2} $ we obtain the viscous Burgers equation 
\begin{equation}
\partial_{\tau} K(\chi) + \partial_{\chi} K(\chi) K(\chi) =\nu \partial_{\chi}^2 K(\chi).
\end{equation}
Solutions of the latter can be obtained via the heat kernel given that the viscous Burgers equation can be mapped to the heat equation via the Cole-Hopf transformation.
To generalize the transform in the matrix case let us consider solutions of the form $K(\chi) = f(\chi) {\bf b}$ where ${\bf b}$  is an ${\cal N} \times {\cal N}$ matrix: 
${\bf b} ^2= \kappa {\bf b}$ and $f(\chi)$ is a scalar, which then satisfies the scalar Burgers equation with rescaled time $\hat \tau = \kappa \tau$ 
and diffusion constant $\hat \nu = {\nu\over \kappa}$.  The scalar Cole-Hopf transformation can now be implemented:
$f(\chi) = -2\hat \nu \partial_{\chi}\big (\log\varphi(\chi)\big )$ where $\varphi$ is a solution of the heat equation: $\partial_{\hat \tau}\varphi(\chi) = \hat \nu\partial^2_{\chi}\varphi(\chi)$.
Interestingly in \cite{DoikouMalhamWiese} it was shown that the quantum discrete NLS model yields in the continuum limit the stochastic heat equation and hence the viscous Burgers equation. 
These are significant connections that are at the epicentre of ongoing investigations at both classical and quantum level (see also relevant findings in \cite{DoikouFindlay, DoikouMalhamWiese2} 
as well as section 4).
\end{itemize}

\noindent 
Some noteworthy comments are in order here. The issue of canonical quantization emerges naturally in this context. In general, the quantities ${\mathbb B}(x,x),\  {\mathbb C}(x,x)$ 
can be seen as the corresponding non-commutative fields, given the relevant definition of the fields (\ref{fields1}).
If one further considers the symmetric case $M=N \to \infty$,  and requires that the fields satisfy canonical commutation relations, 
then one indeed recovers the quantum version of the NLS model with an underlying $\mathfrak{sl}_2$ algebra (see relevant results on the quantum GLM equation for the NLS model in \cite{Thucker}). 
The notion of the non-commutative/quantum Riccati equations associated to the system is thus a particularly pertinent issue, which will be discussed in section 4.
In fact, the use of the Riccati  equation is fundamental in solving the auxiliary problem and deriving the associated non-commutative conserved quantities (see also \cite{DoikouFindlay} for relevant findings).

\section{Dressing transformations as differential operators}

\noindent 
We have discussed in the preceding section the dressing formulation employing an integral operator 
as the Darboux-dressing transformation. We now turn our attention to 
the case where the Darboux transform is a differential operator. The fundamental transform is expressed as a first order 
differential operator:
\begin{equation}
{\mathbb G} = {\mathbb I} \partial_x + K(x), \label{gg1}
\end{equation}
where 
\begin{equation}
 K(x)=\begin{pmatrix}
  {\mathrm A}_{N\times N}(x,y)& {\mathrm B}_{N\times M}(x,y)\cr
 {\mathrm C}_{M\times N}(x,y)& {\mathrm D}_{M\times M}(x,y)
\end{pmatrix}. \label{GrasK}
\end{equation}
Let us now solve the set of dressing relations:
\begin{eqnarray}
&& \Big ({\mathbb I} \partial_x + K(x)\Big){\cal W}\partial_x = \Big ({\cal W}\partial_x +U(x)\Big)\Big ({\mathbb I} \partial_x + K(x)\Big), \label{one}\\
&& \Big ( {\mathbb I} \partial_x + K(x)\Big) \Big  ({\mathbb I} \partial_{t_1} - \hat {\cal W} \partial_x \Big)= \Big  ({\mathbb I} \partial_{t_1} - \hat {\cal W} 
\partial_x + a(x) \Big)\Big ( {\mathbb I} \partial_x + K(x)\Big) , \label{twoa}\\
&& \Big ( {\mathbb I} \partial_x + K(x)\Big) \Big  ({\mathbb I}\partial_{t_n} - {\mathbb I}\partial_x^n\Big) =\Big ({\mathbb I} \partial_{t_n} - {\mathbb I} 
\partial_x^n +\sum_{k=0}^{n-1} a_k(x)\partial_x^k\Big  ) \Big ({\mathbb I} \partial_x + K(x)\Big), n>1.\label{two}
\end{eqnarray}
Equation (\ref{one}) provides:
\begin{eqnarray}
&& U(x)K(x) =- {\cal W}\partial_x K(x), \nonumber\\
&& U(x) =K(x){\cal W} - {\cal W} K(x),
\end{eqnarray}
and as in the integral case discussed above $U$ is defined in (\ref{UU}), where the fields are now identified as
\begin{equation}
\hat u(x) =- h{\mathrm B}(x,x), ~~~~u(x) = h{\mathrm C}(x,x), ~~~\mbox{and} ~~~ h = w_1 -w_2.
\end{equation}
Also, the following relations emerge, similar to the ones appearing  in the integral case (\ref{constr1})
\begin{eqnarray}
&& w_1\partial_x {\mathrm A}(x) = h{\mathrm B}(x){\mathrm C}(x), 
~~~~w_2\partial_x {\mathrm D}(x) = -h{\mathrm C}(x){\mathrm B}(x), \nonumber\\
&& w_1 \partial_x {\mathrm B}(x) = h{\mathrm B}(x){\mathrm D}(x),
 ~~~~w_2 \partial_x {\mathrm C}(x)  =- h{\mathrm C}(x){\mathrm A}(x). \label{constr2}
\end{eqnarray}
From the analysis above we conclude that the ${\mathbb L}^{(0)}$ operator reads as
\begin{eqnarray}
&&{\mathbb L}^{(0)} ={\cal W} \partial_x +  \begin{pmatrix}
0& \hat u \\
u & 0\end{pmatrix}.
\end{eqnarray}

The time dependent part of the dressing (\ref{twoa}), (\ref{two}) also gives sets of equations associated to each time flow  $t_n$:

\begin{itemize}

\item ${\bf n=1}$: equation (\ref{twoa}) yields: $a(x) =-\kappa U(x)$, where $\kappa = {\hat w_1 -\hat w_2 \over w_1 - w_2}$, 
and 
\begin{equation}
\partial_{t_1}K(x)=\hat {\cal W} \partial_x K(x) + \kappa U(x) K(x) .
\end{equation}

\item ${\bf n>1}$: equation (\ref{two}) gives
\begin{eqnarray}
\partial_{t_n} K(x)- \partial_x^n K(x)  + \sum_{k=0}^{n-1} a_k(x) \partial_x^k K(x)  =0,
\end{eqnarray}
as well as the contributions proportional to $\partial_x^m$, which essentially determine each one of the factors $a_k$ of every ${\mathbb L}^{(n)}$ operator
\begin{equation}
\sum_{k=1}^n a_{k-1}(x)\partial_x^k- \sum_{m=1}^{n-1} \begin{pmatrix}
  n\cr
 m
\end{pmatrix} \partial_x^{n-m}K(x) \partial_x^{m} + \sum_{k=0}^{n-1}\sum_{m=1}^{k-1} \begin{pmatrix}
 k\cr
 m
\end{pmatrix}   a_k(x)\partial_x^{k-m}K(x)\partial_x^m =0.
\end{equation}
The latter relation gives rise to a tower of constraints among the various coefficients $a_k$, which can then be uniquely determined. 
Moreover,  use of the extra constraint provided by (\ref{one}) gives a solution (one soliton solution) of the underlying non-linear PDE. 
\end{itemize}

We report below the Lax pairs ${\mathbb L}^{(n)}$,  as well as the relevant equations of motion for $n =1,\  2, \ 3,$ i.e. 
for the generalized transport, NLS and mKdV equations  respectively. 
\begin{itemize}

\item {\tt ${\bf  n=1}$: the matrix transport equation} 

The time component for the $t_1$ time flow reads as
\begin{eqnarray}
&& {\mathbb L}^{(1)} = {\mathbb I} \partial_{t_1}- \hat {\cal W} \partial_x - \kappa \begin{pmatrix}
0& \hat u \\
u & 0\end{pmatrix}.
\end{eqnarray}
The zero curvature condition (commutativity) for the pair ${\mathbb L}^{(0)},\  {\mathbb L}^{(1)},$ leads to the matrix transport equation:
\begin{equation}
\partial_{t_1} U = \big (\hat {\cal W} - \kappa {\cal W} \big) \partial_x U,
\end{equation}
where $U =  \begin{pmatrix}
0& \hat u \\
u & 0\end{pmatrix}$.

\item {\tt ${\bf n=2}$: the matrix NLS equation} 

The time part of the Lax pair for the NLS model $n=2$ is given by
\begin{eqnarray}
&& {\mathbb L}^{(2)} = {\mathbb I} \big (\partial_{t_2} -\partial_x^2\big )+{2\over h} \begin{pmatrix}
-{1\over w_1}\hat u u & - \partial_x\hat u \\
 \partial_x u & {1\over w_2}u \hat u \end{pmatrix}.
\end{eqnarray}
Consequently the equations of motion for NLS read as
\begin{eqnarray}
&& \partial_{t_2} u -{w_1 +w_2 \over h } \partial_x^2   u+{2(w_1+w_2)\over h w_1 w_2 }u \hat u u =0, \label{matrixem}
\end{eqnarray}
similarly for $\hat u$, but $t_2 \to - t_2$.  

\item {\tt ${\bf n=3}$: the generalized  mKdV model}

In this case the time component of the Lax pair is
\begin{eqnarray}
 {\mathbb L}^{(3)} &=& {\mathbb I}\big (\partial_{t_3} - \partial_x^3\big  )+ {3\over h} \begin{pmatrix}
-{1\over w_1}\hat u u & - \partial_x\hat u \\
 \partial_x u & {1\over w_2}u \hat u \end{pmatrix}\partial_x  \nonumber\\ &+& {3\over h^2}\begin{pmatrix}
-\hat u \partial_x u + {w_1\over w_2 }\partial_x\hat u u & w_2 \partial^2_x\hat u-{w_1+w_2\over w_1w_2}\hat u u 
\hat u\\
w_1 \partial^2_x u-{w_1+w_2\over w_1w_2} u \hat u  u  & \partial_x  u \hat u -{w_2\over w_1 }u \partial_x\hat u   \end{pmatrix}. \nonumber\\
&& 
\end{eqnarray}
It is worth noting that for the derivation of the generalized matrix NLS and mKdV ${\mathbb L}^{(n)},\ ( n =2,\ 3)$ operators the use of the constraints (\ref{constr1}), (\ref{constr2}) 
in both the integral and differential cases has been essential. 
The equations of motion for the generalized mKdV then read as
\begin{eqnarray}
&& \partial_{t_3} u -{{\mathbb E}\over h^2 } \partial_x^3   u+{3{\mathbb E}\over h^2 w_1 w_2}(u \hat u \partial_x  u + \partial_x u \hat u u) =0,
\end{eqnarray}
where ${\mathbb E} = w_1^2 + w_2^2 + w_1 w_2$, and  similarly for $\hat u$.

\item {\tt Matrix Burgers equation}\\
As discussed in the integral case to obtain the matrix Burgers equation we
focus on the bare operator ${\mathbb I}(\partial_t -\partial_x^2) $ and dressed operator ${\mathbb I}(\partial_t -\partial_x^2) +V(x)$.
We consider the first order differential operator ${\mathbb I}\partial_x + K(x)$, ($V,\ K$ are ${\cal N} \times {\cal N}$ matrices), as the fundamental Darboux transformation.
Then via the dressing process as described previously in this subsection we conclude that $V(x) = 2\partial_x K(x)$ and:
\begin{equation}
\partial_{\tau}K(x) -{1\over 2}\partial^2_x K(x) + \partial_x K(x) K(x) =0,
\end{equation}
where $\tau = 2 t$. The equation we obtain now is more restrictive compared to the one we got in the integral case. Now the diffusion constant is fixed, 
so we can only deal with the viscous Burgers equation as opposed to the integral case where both viscous and inviscid Burgers equations emerge.

\end{itemize}

\subsection{General Darboux transform $\&$ solutions}

\noindent 
Consider now the most general scenario for the Darboux transform given as an $m^{th}$ order differential operator. 
We focus here on the time independent part of the transform and find the
tower of differential equations obeyed by the coefficients. Let
\begin{equation}
{\mathbb G} = {\mathbb I}\partial^m_x + \sum_{k=0}^{m-1} b_k(x) \partial_x^k. \label{GG0}
\end{equation}
Our aim is to identify the coefficients $b_k$ of the differential operator. Indeed, by means of the fundamental dressing relation
\begin{equation}
\Big ({\mathbb I} \partial_x^m + \sum_{k=0}^{m-1} b_k(x) \partial_x^k\Big ) {\cal W} \partial_x =\Big  ({\cal W} \partial_x + U(x)\Big ) 
\Big ({\mathbb I}\partial_x^m + \sum_{k=0}^{m-1} b_k(x)\partial_x^k\Big ),
\end{equation}
we find recursion relations for the matrix coefficients $b_k$:
\begin{eqnarray}
&& {\cal W}b_{k-1} - b_{k-1}{\cal W} + {\cal W} \partial_x b_k + Ub_k =0, ~~~~k \in \{1, \ldots, m-1 \} \nonumber\\
&&  Ub_0 + {\cal W} \partial_x b_0 =0,\nonumber\\
&& {\cal W} b_{m-1} - b_{m-1}{\cal W} +U =0, \label{gen-con}
\end{eqnarray}
which yield  the form of $U$ and $b_k$ as well as solutions of the non-linear equations. 

Note that in earlier works \cite{Degalomba1, Degalomba2, ADP} the Lax pairs as well as the general Darboux transforms for the vector and matrix NLS models were expressed 
as $c$-number matrices. Here we are offering a unifying dressing scheme regardless of the particular form of the operators. We have already examined the case of integral and 
differential operators and we have seen their equivalence, while in section 4.1 we present the matrix dressing process, which is naturally in one to one correspondence with the dressing 
when considering the Darboux to be a differential operator.

It will also be instructive to comment on the distinct choices of Darboux-dressing transforms considered so far i.e. the integral Darboux versus the differential one. 
Indeed, the main advantage  when considering  the integral Darboux transform  is that solutions of the GLM equation found via the linear data are provided
in a straightforward way yielding in turn generic solutions of the associated integrable nonlinear PDEs.  In the case of the differential Darboux 
transform on the other hand  -- as is the case of the matrix Darboux studied in section 4 -- the dressing  of the linear operators  is a much simpler 
process given that one does not  have to deal with the  involved boundary terms emerging when performing the numerous integrations by parts in the integral case. 

In the preceding subsection a detailed account on the dressing of the linear operators and the construction of the integrable NLS hierarchy was provided for the 
fundamental Darboux, whereas in this section we deal with the generic differential Darboux and obtain the fundamental dressing relations for the 
$t$-independent part of the Lax pairs (\ref{gen-con}).
We restrict our attention to the construction of solutions via the set of time constraints (\ref{gen-con}).
Let us for simplicity focus on the one soliton case here to illustrate the process, a more exhaustive analysis on generic solutions via the recursion relations (\ref{gen-con}) 
will be presented in a forthcoming publication. It will be important here as is the case in the usual Darboux transformation for the familiar NLS to 
consider the following ansatz for the one soliton solution, in accordance also with the results of the previous subsection:
\begin{equation}
{\mathrm A}(x) = {\tt A}(x)b \hat b,  ~~~{\mathrm D}(x) = {\tt D}(x)\hat b b,  ~~~{\mathrm B}(x) = {\tt B}(x)b,  ~~~{\mathrm C}(x) = {\tt C}(x)\hat b,  \label{ansatz2}
\end{equation}
where $b,\hat b$ are $N\times M$ and $M\times N$ matrices.
We also require: ${\tt B} {\tt C}  = k_1{\tt  A} -\xi_1 {\tt A}^2$. A similar requirement  is also considered for ${\tt D}$ with parameters $k_2,\ \xi_2$.
Note that the ansatz (\ref{ansatz2}) as well as the constraints on ${\tt A},\  {\tt D}$  are compatible with (\ref{constr2}). After  substituting  the expressions (\ref{ansatz2}) 
into the first of the equations  (\ref{constr2}) we obtain 
\begin{equation}
\partial_x {\tt A}(x) =  {h\xi_1 \over w_1}\big  ({ k_1\over \xi_1} {\tt A}(x) -{\tt A}^2(x)\big )\  \Rightarrow\  
{\tt A}(x) = -{k_1\over \xi_1} { e^{-{hk_1\over w_1} (x- x_0)}\over 1 - e^{-{ hk_1\over w_1} (x- x_0)}}.
\end{equation}
Then from equation (\ref{constr2}) we obtain a solution for ${\tt C}$
\begin{equation}
\partial_x {\tt C}(x) = -{h \over w_2} {\tt C}(x) {\tt A}(x) \  \Rightarrow\   {\tt C}(x) = {{\tt C}_0 \over 1- e^{-{ hk_1\over w_1} (x- x_0)}},
\end{equation}
where we have assumed ${w_1\over w_2 \xi_1} =1$. Similar solutions are obtained for the pair ${\tt D},\  {\tt B}$, but are omitted here for brevity. 
Obtaining solutions associated  to more involved Darboux transformations becomes a highly involved algebraic problem given that one has to solve all 
the associated constraints (\ref{gen-con}). This is a very interesting direction to pursue and will be further discussed in future works.


\section{Non-commutative Riccati flows}

\noindent 
We have thus far focused on the derivation of  solutions of the integrable non-linear PDEs and the construction of the matrix NLS hierarchy via the dressing transform.
We have not however discussed the issue of conserved quantities associated to the hierarchy, and this is in fact one of the fundamental objectives of this section.
More precisely, the purpose of this section is two-fold:
1) We derive  non-commutative Riccati flows associated to the matrix NLS hierarchy. 
As mentioned earlier in the text the notion of the non-commutative Riccati  equation is fundamental in solving the auxiliary problem and deriving 
the related non-commutative conserved quantities.
Moreover, we identify the Poisson structure for the classical fields (components of the matrix fields) via the comparison of the equations of motion 
from the zero curvature condition and the Hamiltonian description.
2) We identify the non-commutative Riccati equations associated to the generic matrix Darboux-dressing transform expressed as a formal series expansion. 
This naturally turns out to be equivalent to the Riccati equations derived for the solution of the $x$-part of the auxiliary linear problem 
(see also \cite{Wadati} for a relevant discussion). The Riccati equations are also derived in the case where the Darboux is chosen to be a 
matrix-integral operator leading to non-local quadratic differential equations.

Before we focus on the non-commutative Riccati equations we first  review the dressing process for the matrix NLS hierarchy when the Lax 
pair is comprised of $c$-number matrices; which will then naturally lead to the Riccati equations. 
Also, the equivalence between the matrix case and the matrix-differential case studied in section 3  is exhibited.
We should note that not only are we able to provide the classical charges of the hierarchy, but also 
identify some conserved quantities in a ``block''  form, which may be thought of as the  precursors  of  the quantum conserved charges.

\subsection{The Darboux matrix $\&$ the hierarchy}
\noindent 
We  briefly review the auxiliary linear problem and the dressing process for a Lax pair consisting of generic $c$-number $d \times d$  matrices $ (U, V) $. 
The Lax pair matrices depend in general on some fields and a spectral parameter, and obey the auxiliary linear problem:
\begin{eqnarray}
	&&\partial_x \Psi(\lambda, x,t) = U(\lambda, x,t) \Psi(x,t),  \nonumber\\
 && \partial_{t_n} \Psi(\lambda, x,t)= V^{(n)}(\lambda, x,t) \Psi(\lambda, x,t). \label{ALP}
\end{eqnarray}
For the matrix  NLS hierarchy the $U$-matrix is given as ($d =M+N$)
\begin{equation}
	U(\lambda, x,t_n) = \lb \begin{matrix}
		\frac{ \lambda}{2}{\mathbb I}_{N \times N} & \hat   u(x, t) \\
		 u(x, t) & -\frac{ \lambda}{2} {\mathbb I}_{M\times M}
	\end{matrix} \rb, \label{eq:NLS_Lax}
\end{equation}
where in general $\hat u,\  u$ are $N\times M$ and $ M \times N$ matrices. 
The $U$-matrix (\ref{eq:NLS_Lax}) will be the starting point in our dressing process, 
whereas all the time components will be explicitly derived.

The first aim is to identify the ``dressed'' quantities  $V^{(n)}$ of the hierarchy.
Let $U_0,\ V_0^{(n)}$ be the ``bare'' Lax pairs:
\begin{equation}
U_0(\lambda) = {\lambda \over 2}\Sigma,~~~~V_0^{(n)}(\lambda)= {\lambda^{n} \over 2}\Sigma,
\end{equation}
where we define
\begin{equation}
\Sigma =  \lb \begin{matrix}
		{\mathbb I}_{N \times N} & 0 \\
		 0 & -{\mathbb I}_{M\times M}
	\end{matrix} \rb.
\end{equation}
Also, the ``dressed'' time components of the Lax pairs can be expressed as formal series expansions
\begin{equation}
V^{(n)}(\lambda, x, t) = {\lambda^n \over 2}\Sigma + \sum_{k=0}^{n-1}\lambda^k w^{(n)}_k(x, t),
\end{equation}
where the quantities $ w^{(n)}_k$ will be identified via the dressing transform.

The Darboux transform ${\mathrm G}$ is applied on the ``bare'' auxiliary function
\begin{equation}
\Psi(\lambda, x, t)= {\mathrm G}(\lambda, x,t )\  \Psi_0(\lambda), \label{Darboux2}
\end{equation}
yielding (see also the analogues (\ref{generalDarboux}), (\ref{Darboux0}))
\begin{equation}
\partial_x{\mathrm G} = U\ {\mathrm G}-  {\mathrm G}\ U^{(0)},
~~~~\partial_{t_n}{\mathrm G} = V^{(n)}\ {\mathrm G} -{\mathrm  G}\  V_0^{(n)}\ . \label{fundam2}
\end{equation}
The fundamental Darboux matrix is chosen to be of the form
\begin{equation}
{\mathrm G}(\lambda, x, t)=  \lb \begin{matrix}
		\lambda {\mathbb I}_{N \times N} + {\cal A}_{N\times N}(x, t) & {\cal B}_{N\times M}(x, t) \\
		{\cal  C}_{M\times N}(x, t) & \lambda{\mathbb I}_{M \times M}  +{\cal D}_{M\times M}(x, t)
	\end{matrix} \rb = \lambda {\mathbb I} + {\cal K}. \label{GG}
\end{equation} 
This is the simplest case to consider, but nevertheless it fully describes the dressing process for the Lax pair. 
By solving the $x$-part of equations (\ref{fundam2}) we obtain
\begin{equation}
\hat u =-{\cal  B}, ~~~u ={\cal C} ~~~\mbox{and} ~~~\partial_{x} {\cal K} =   \lb \begin{matrix}
		0 & \hat u\\
		u &0
	\end{matrix} \rb {\cal K}. \label{constr2b}
\end{equation}
Note that in all the expressions below $x$ and $t$ dependence is implied.

From the time part of (\ref{fundam2}) we obtain a set of recursion relations, in exact analogy to the case of dressing via a differential operator, i.e.
\begin{eqnarray}
&& w^{(n)}_{n-1}= {1\over 2}\big [{\cal K},\ \Sigma\big ], \nonumber \\
&& w^{(n)}_{k-1} =-w^{(n)}_{k}\  {\cal K}, ~~~~k\in\{1, 2, \ldots, n-1\}\nonumber\\
&& \partial_{t_n} {\cal K} = w^{(n)}_0\  {\cal K}. \label{recursion1}
\end{eqnarray}
We solve the latter recursion relations, and identify the first few time components of the Lax pairs:
\begin{eqnarray}
&& V^{(0)} = {1\over 2} \Sigma, \nonumber\\
&& V^{(1)} = \lambda V^{(0)} +  \lb \begin{matrix}
		0 & \hat u\\
		u &0
	\end{matrix} \rb, \nonumber\\
&& V^{(2)}=   \lambda V^{(1)} + \lb \begin{matrix}
		-\hat u u & \partial_x\hat u\\
		-\partial_x u & u \hat u
	\end{matrix}\rb,\nonumber\\
&& V^{(3)}=  \lambda V^{(2)} +  \lb \begin{matrix}
		\hat u\ \partial_xu - \partial_x\hat u\  u& -2 \hat u u \hat u +\partial_x^2 \hat u\\
		-2 u \hat u u + \partial_x^2 u  &u\  \partial_x \hat u - \partial_x u\ \hat u
	\end{matrix} \rb.
\end{eqnarray}
In general, the $V^{(n)}$ operator is  identified  as $V^{(n)} = \lambda V^{(n-1)} + w_0^{(n)}$. 
 Moreover, the recursion relations (\ref{recursion1})  lead to
\begin{equation}
w_k^{(n)} = w_{k-1}^{(n-1)}, ~~~~~w_{0}^{(n)} = (-1)^{n-1} w_{0}^{(1)}{\cal K}^{n-1},
\end{equation}
where the latter relations together with the constraints (\ref{constr2b}) suffice to provide $w_0^{(n)}$ at each order\footnote{Note that in the special case 
$\hat u = u$ ($N=M$) the Lax pair $(U,\ V^{(3)})$ is the one of the matrix mKdV model,  see also \cite{AKNS1, Clark} for the Lax pair of the usual mKdV model.}. 

Note that (\ref{fundam2}), (\ref{GG}) also produce the fundamental Darboux-B\"acklund transformation (BT) that connects two 
different solutions of the underlying integrable PDEs, i.e the pair $U,\  V^{(n)}$ are associated to the fields $u,\ \hat u$, 
whereas $U_0,\  V^{(n)}_0$ are associated to the fields $u_0,\  \hat u_0$. Then the $x$-part 
of (\ref{fundam2}) leads to the matrix Darboux-BT relations:
\begin{eqnarray}
&& {\cal B}= -(\hat u - \hat u_0), ~~~~~{\cal C}= u - u_0, \nonumber\\
&& \partial_x{\cal A} =\hat u {\cal C} - {\cal B}u_0 , ~~~~\partial_x {\cal D} =   u {\cal B}- {\cal C}\hat u_0, \nonumber\\
&& \partial_x  {\cal B} = \hat u{\cal  D}- {\cal A} \hat u_0 , ~~~~ \partial_x {\cal C} = u {\cal A}-  {\cal D} u_0.
\end{eqnarray}
For the dressing process we have clearly used the trivial solution $u_0 = \hat u_0 =0$.

Let us also briefly discuss the general Darboux expressed as a formal $\lambda$-series expansion
\begin{equation}
{\mathrm G} = \lambda^m {\mathbb I}+ \sum_{k=0}^{m-1} \lambda^k g_k, \label{generalg1}
\end{equation}
where $g_{k}$ are ${\cal N} \times {\cal N}$ matrices to be identified.
We focus on the fundamental recursion relations arising from the $x$-part of the Darboux transform (\ref{fundam2}):
\begin{eqnarray}
&&  \lb \begin{matrix}
		0 & \hat u\\
		u &0
	\end{matrix} \rb= {1\over 2} \big [g_{m-1},\ \Sigma \big ],  \label{c1}\\
&&\partial_x g_0 =   \lb \begin{matrix}
		0 & \hat u\\
		u &0
	\end{matrix} \rb\ g_0, ~~~~~ \partial_x g_k = {1\over 2} \big [ \Sigma,\ g_{k-1} \big ]+  \lb \begin{matrix}
		0 & \hat u\\
		u &0
	\end{matrix} \rb\ g_k. \label{c2}
\end{eqnarray}
The infinite series expansion $m \to \infty$  is particularly interesting (see also e.g. \cite{loop1, Uhlenbeck} in relation  to the  loop group action), and discussed in subsection 4.3, 
where the non-commutative Riccati equation associated to the generic Darboux transform is identified.

Some general comments are in order here.
It is clear that the choice of ${\mathrm G}$ is compatible with the generic dressing constraints (\ref{generalDarboux}).
More precisely, for the dressing process as described in sections 3 and 4:  we consider the {\it fundamental  Draboux} given
by a first order matrix differential operator (\ref{gg1}), (\ref{GrasK}) or equivalently by  (\ref{GG}) in the matrix realization. 
Analogously, the dressing equations (\ref{one})--(\ref{two}) are equivalent to (\ref{fundam2}). Having chosen the form of
the fundamental Darboux we can uniquely derive it via the dressing relations (\ref{one})--(\ref{two})  or (\ref{fundam2}), and
moreover we can find the respective solution of the associated integrable PDEs, as discussed in subsection 3.1.  
Naturally, the general transform (\ref{GG0}) is also equivalent to (\ref{generalg1}). The case $m \to \infty$ (\ref{generalg1}) 
as a formal power series expansion is discussed together with the integral representation in the last subsection when deriving the associated Riccati equations.

Note that we could have chosen another fundamental Darboux e.g. a differential-matrix operator of the form ${\mathrm G} = {\mathrm D} \partial_x + {\mathrm K}$ or a matrix 
${\mathrm G}= \lambda {\mathrm D} + {\mathrm K}$, where ${\mathrm D}$ is a constant ${\cal N} \times {\cal N}$ matrix that commutes 
with the bare operators and ${\mathrm K}$ is of the generic form (\ref{GrasK}). This is equivalent to our chosen fundamental Darboux up to a gauge. 
When ${\mathrm D}$ is singular e.g. ${\mathrm D}  = \mbox{diag}({\mathbb I}_{N\times N},\ 0_{M\times M})$ one obtains in principal a different set of dressing relations 
from (\ref{one})--(\ref{two}) or (\ref{fundam2}), but we are not examining this case here.

\subsection{The non-commutative Riccati equations}

\noindent 
We come now to our main task, which is the study of the non-commutative Riccati equations for the matrix NLS hierarchy.
The primary aim is to identify solutions of the auxiliary linear problem and hence classical and non-commutative conserved quantities. 
The auxiliary function is expressed as $\Psi =  \lb \begin{matrix}
		\Psi_1 \\
		\Psi_2
	\end{matrix} \rb$,
where we choose $\Psi_1$ to be an $N\times N$ matrix and consequently $\Psi_2$ an $M\times N$ matrix. 
We could have chosen equivalently $\Psi_1,\  \Psi_2$ to be $N\times M$ and $M \times M$ matrices respectively.
We proceed essentially as in the scalar case (see e.g. \cite{FT}), noting that the non-commutativity between 
the components $\Psi_1,\ \Psi_2$ leads to a non-trivial behavior when explicitly solving the associated Riccati equation.
Consider now the $x$-part of the auxiliary problem (\ref{ALP}):
\begin{eqnarray}
&& \partial_x\Psi_1 = {\lambda \over 2}\Psi_1 +\hat u \Psi_2, \nonumber\\
&&   \partial_x\Psi_2 = -{\lambda \over 2}\Psi_2 + u \Psi_1.
\end{eqnarray}
Define the $M\times N$ matrix $\Gamma = \Psi_2\ \Psi_1^{-1}$, then from the latter expressions 
one arrives at the non-commutative Riccati equation obeyed by $\Gamma$:
\begin{equation}
\partial_x \Gamma = u -\lambda\Gamma -\Gamma\ \hat u\  \Gamma. \label{Ric}
\end{equation}
The next important task is to identify the element $\Gamma$. This can be achieved as in the usual NLS case \cite{FT}
by expressing $\Gamma$ in a formal power series expansion $\Gamma = \sum_{k} {\Gamma^{(k)} \over \lambda^k}$ 
and solve the Riccati equation at each order.
Then (\ref{Ric}) reduces to: $\Gamma^{(1)} = u$ and
\begin{eqnarray}
\partial_x \Gamma^{(k)}= - \Gamma^{(k+1)} - \sum_{l=1}^{k-1}\Gamma^{(l)}\  \hat u\  \Gamma^{(k-l)}, ~~~~k >0.
\end{eqnarray}
Let us report below the first few terms of the expansion
\begin{eqnarray}
&& \Gamma^{(1)}= u , ~~~~~~\Gamma^{(2)}=-\partial_x u, ~~~~~~ \Gamma^{(3)}=\partial_x^2u -u\hat u u,~~\ldots \label{gammas}
\end{eqnarray}

Some particularly relevant findings on Grassmannian/Riccati flows are presented 
in \cite{BDMS, BDMS2}, where the infinite Grassmannian was exploited in order to produce non-local PDEs.
Here on the other hand we deal with the finite Grassmannian $Gr(N|N+M)$, although 
at the quantum level we are interested in the infinite limit of $Gr(N|2N)$ as $N\to \infty$.
A more exhaustive investigation on the significant notion of non-commutative Riccati flows 
as well as comparison with the relevant findings at the discrete quantum level \cite{DoikouFindlay} 
will be presented in forthcoming works.

The next natural step is to identify the associated conserved quantities by means of the  the auxiliary linear problem relations.
Let us express the time components of the Lax pairs as $V^{(n)} =    \lb \begin{matrix}
		\alpha_n&  \beta_n\\
		\gamma_n &\delta_n
	\end{matrix} \rb$, and also recall for both the $x$ and $t$-parts of the linear problem:
\begin{eqnarray}
&& \partial_x \Psi_1\ \Psi_1^{-1} =  {\lambda \over 2} + \hat u \Gamma,   \nonumber\\
&&  \partial_{t_n} \Psi_1\ \Psi_1^{-1} = \alpha_n +\beta_n \Gamma.
\end{eqnarray}
After cross differentiating the equations above we conclude
\begin{equation}
\partial_{t_n}\big (\hat u \Gamma\big ) = \partial_x\big (\alpha_n + \beta_n \Gamma\big ) + \big [ \alpha_n + \beta_n \Gamma,\  \hat u \Gamma \big ]. \label{basis}
\end{equation}
In the scalar (commutative) NLS case the commutator in the latter equation is zero, and thus the quantities 
\begin{equation}
I^{(k)} = \int_{\mathbb R} dx\ \hat u(x) \Gamma^{(k)}(x), \label{block}
\end{equation}
are automatically conserved (see also \cite{FT}); where we have assumed vanishing boundary conditions at $\pm \infty$.
In the non-commutative case however the commutator in (\ref{basis}) is in principle non-zero, except when certain constraints are imposed. 
Let us focus on the first two simple members of the hierarchy i.e. $V^{(0)},\  V^{(1)}$ by just replacing the corresponding $\alpha_n,\ \beta_n,\  n=0,\ 1$ 
we can immediately see that the commutator is zero.  Of course we are  interested in the case $M=N $ and more precisely in the thermodynamic limit $N\to \infty$. 

Indeed, let us focus on the symmetric case $M =N$ and in the limit $N\to \infty$.
It is important to note that to systematically address the issue of conservation laws in the non-commutative and in particular the quantum cases 
we need to assume some kind of ultra-locality, especially if we also wish to make a direct connection with the discrete quantum case. 
The obvious ultra-locality condition is of the form 
\begin{equation}
\big [ \partial_xu(x),\ u(x)\big ] =0, \label{ultra}
\end{equation} 
which naturally reflects the fact,  in the discrete quantum set up,  that any two matrices $A,\  B$ acting on different ``quantum'' spaces commute, i.e.
$A_1 B_2 = B_2 A_1$\footnote{Here we use the standard index notation, for any $d\times d$ matrix $A$:
\begin{equation}
A_1 = A \otimes {\mathbb I}, ~~~~A_2 = {\mathbb I} \otimes A, \label{notation1}
\end{equation}
where ${\mathbb I}$ is the $ d \times d$ identity matrix.},
where in principle $A _{\alpha} B_{\alpha} \neq B_{\alpha} A_{\alpha}$.  The consistent quantum continuum limit should lead to 
(\ref{ultra}) (see also a very relevant description at the classical level in \cite{AvanDoikouSfetsos}).
These ultra-locality conditions should also be compatible  with the quantum involution of the charges $I^{(k)},$ i.e. 
compatible with a quantum Hamiltonian formulation. We shall report on the important subject of block conserved quantities, 
and in particular in relation to the issue of ultra-locality  in detail in a separate publication.

Note also that in the continuum limit $N \to \infty$, the matrix operators 
$\Gamma,\ u, \hat u,\ \alpha,\ \beta$ with elements $\Gamma_{ij},\ u_{ij},\ldots$ (we have suppressed the subscript $_n$ for simplicity) 
turn to integral operators with kernels $\Gamma(\xi,\eta),\ u(\xi, \eta), \ldots$.
It is clear that by taking the trace in both cases (discrete and continuous) we eliminate the unwanted commutator in (\ref{basis}), 
and we obtain the following conserved quantities:
\begin{eqnarray}
&& {\cal I}^{(k)} = \int_{\mathbb R} dx\ \sum_{i,j=1}^N \hat u_{ij}(x) \Gamma^{(k)}_{ji}(x)  ~~~ \mbox{Discrete case}, \label{conserved1}\\
&&  {\cal I}^{(k)} = \int_{\mathbb R} dx\ \int_{{\mathrm I}}\ d\xi \int_{{\mathrm I}}\ d\eta\  \hat u(x; \xi, \eta)\Gamma^{(k)}(x; \eta, \xi) 
~~~~\mbox{Continuous case}, 
\end{eqnarray}
where the interval of integration ${\mathrm I}$ can be in general ${\mathbb R}$.
The conserved quantities above have been derived independently of the existence of ultra-locality conditions, 
however these are scalar objects as opposed to the block (potentially quantum) objects (\ref{block}), i.e. these are the classical conserved quantities. 

We are now  focusing  on $k=2,\ k=3$, which correspond to the matrix transfer equation and the matrix NLS equation respectively. 
Then the conserved quantities become (\ref{conserved1}), after also recalling the expressions for $\Gamma^{(2)},\ \Gamma^{(3)}$ (\ref{gammas})
\begin{eqnarray}
&& {\cal I}^{(2)} = -  \sum_{i, j}\int_{\mathbb R}dx\  \hat u_{ij} \partial_x u_{ji}\nonumber\\
&& {\cal I}^{(3)} =  \sum_{i, j}\int_{\mathbb R}dx\  \hat u_{ij}\Big (\partial_x^2 u_{ji} - \sum_{m, n} u_{jm}\hat u_{mn} u_{ni} \Big ).
\end{eqnarray}
Moreover, the equations of motion from the zero curvature conditions for the Lax pair $(U,\ V^{(k)}),\ k =\{2,\ 3\}$ are generalized transfer and NLS-type equations, 
expressed in the compact form (see also (\ref{matrixem}))
\begin{eqnarray}
&& \partial_{t_2} u -\partial_x u =0 ~~~~~~~~~~~~~~~\mbox{k=2, matrix transfer equation}  \nonumber\\
&& \partial_{t_3} u + \partial_x^2 u - 2 u \hat u u =0 ~~~~~\mbox{k=3, matrix NLS equation}. \label{matrixem2}
\end{eqnarray}
Now considering the charges ${\cal I}^{(k)}$ as Hamiltonians and using the following Poisson structure for the components $u_{ij},\ \hat u_{ij}$
\begin{equation}
\Big \{ \hat u_{ij}(x),\ u_{mn}(y)\Big \} = \delta_{in} \delta_{mj} \delta(x- y), \label{Poisson}
\end{equation}
we recover the equations of motion (\ref{matrixem2}) via: $\partial_{t_n} u_{kl} =\big \{ {\cal H}^{(n)},\ u_{kl}\big \}$ (where ${\cal H}^{(2)} = -{\cal I}^{(2)},\  
{\cal H}^{(3)} = {\cal I}^{(3)}$), a fact that confirms the validity of (\ref{Poisson}), and also guarantees the involution of the charges.

\subsection{Riccati equations for the general Darboux transform}
\noindent
In this subsection we work out the Riccati equations associated  to the generic Darboux-dressing transform. 
This will be achieved for both cases considered in this investigation i.e. when the Lax pairs are given in terms of matrices 
and when they are given in terms of matrix-differential operators

Let us first consider the Lax pair as well as the Darboux transform expressed in terms of matrices, which is basically the case we address in this section.
We express the transform in the most general form as a formal power series expansion (see also (\ref{generalg1}) $ m\to \infty$):
\begin{equation}
{\mathrm G}(\lambda,x,t) = \sum_{k=0}^{\infty} {{\mathfrak g}_k(x,t) \over \lambda^k},
\end{equation}
where we consider ${\mathfrak g}_0= {\mathbb I}$, and this choice is naturally compatible with equations (\ref{fundam2}) satisfied by ${\mathrm G}$.
Let also 
\begin{equation}
{\mathrm G}(\lambda, x, t)=  \lb \begin{matrix}
		 {\mathrm A}_{N\times N}(\lambda, x, t) & {\mathrm B}_{N\times M}(\lambda, x, t) \\
		{\mathrm  C}_{M\times N}(\lambda, x, t) & {\mathrm D}_{M\times M}(\lambda, x, t)
	\end{matrix} \rb. \label{GG2}
\end{equation}
Substituting the latter to the $x$-part of (\ref{fundam2}) and focusing on the first column of 
the matrix i.e. on the elements ${\mathrm A}$ and ${\mathrm C}$:
\begin{eqnarray}
&& \partial_x{\mathrm A} = \hat u {\mathrm C} \nonumber\\
&& \partial_x {\mathrm C} = -\lambda {\mathrm C} + u {\mathrm A}.
\end{eqnarray}
After we define $\Gamma = {\mathrm C} {\mathrm A}^{-1}$, we 
conclude that $\Gamma$ satisfies the Riccati equation (\ref{Ric}), derived from the solution of the linear auxiliary problem in the preceding subsection.
The solution of the Riccati equation and hence the derivation  of the entries of the Darboux matrix is achieved via the series expansion 
of $\Gamma$ in powers of $\lambda^{-1}$ as discussed in subsection 4.2.
Equivalently we could  have focused on the elements ${\mathrm B}$ and ${\mathrm D}$, and define $\hat \Gamma = {\mathrm B} {\mathrm D}^{-1}$, 
and hence obtain an analogous Riccati equation for $\hat \Gamma$:
\begin{equation}
\partial_x \hat \Gamma = \hat u +\lambda\hat \Gamma - \hat \Gamma u \hat \Gamma. \label{Ric2}
\end{equation}
The equivalence between the Riccati equations associated to the solution of the auxiliary linear problem as discussed is the previous section, 
and the Riccati equation derived from the Darboux transform above is obvious.
It is also worth noting that even in the case of the fundamental transform  
a quadratic equation for the transform emerges naturally, 
and provides a solution to the non-linear integrable equation as discussed in section 4.1, (see also equivalent findings in subsection 3.1). 

We come now to the case  where the Lax pair is given in terms of matrix-differential operators and the Darboux transform given as matrix-integral operator, 
which is the case discussed in section 2. We consider the generalized equations (\ref{generalDarboux}) and focus on the $n=0$ case 
\begin{equation}
{\mathbb G}\ A^{(0)} = {\mathbb A}^{(0)}\ {\mathbb G}. \label{dres0}
\end{equation}
Now the Darboux transform ${\mathbb G}$ is expressed as a matrix with entries being integral operators, and the latter equation is the equivalent 
of the $x$-part of (\ref{fundam2}). For our particular example here
\begin{equation}
{\mathbb A}^{(0)} = \lb \begin{matrix}
		{\mathrm X}& \hat u\\
		u  & {\mathrm Y}
	\end{matrix} \rb, ~~~~~A^{(0)} = \lb \begin{matrix}
		{\mathrm X}& 0\\
		0  & {\mathrm Y}
	\end{matrix} \rb,
\end{equation}
where we define
\begin{equation}
{\mathrm X} =w_1{\mathbb I}_{N\times N} \partial_x,~~~~~~ 
{\mathrm Y} = w_2{\mathbb I}_{M\times M}  \partial_x, ~~~~~~{\mathbb G} = \lb \begin{matrix}
		{\mathfrak A}_{N\times N}& {\mathfrak B}_{N\times M}\\
		{\mathfrak C}_{M\times N} & {\mathfrak D}_{M\times M}
	\end{matrix} \rb \label{def}
	\end{equation}	
and the entries of ${\mathbb G}$ are matrix-integral operators. We focus on the elements of the first column in  (\ref{dres0}) and derive for the elements ${\mathfrak A}$ and ${\mathfrak C}$:
\begin{eqnarray}
&& {\mathfrak A} {\mathrm X} ={\mathrm X}{\mathfrak A} + \hat u {\mathfrak C},  \nonumber\\ 
&& {\mathfrak C}{\mathrm X}  = u{\mathfrak A} + {\mathrm Y} {\mathfrak C}. 
\end{eqnarray}
We then define ${\mathfrak G} = {\mathfrak C}\ {\mathfrak A}^{-1}$ (provided that ${\mathfrak A},\ {\mathfrak D}$ are invertible), and obtain the quadratic (Riccati) equation for ${\mathrm G}$
\begin{equation}
{\mathfrak G} {\mathrm X}- {\mathrm Y}{\mathfrak G}= u -{\mathfrak G} \hat u{\mathfrak G}. \label{ricc1}
\end{equation}
${\mathfrak G}$ is an integral operator with kernel $\gamma(x,z)$, hence the integral representation of (\ref{ricc1}) leads to (see also (\ref{def}), and recall $h= (w_1-w_2)$)
\begin{eqnarray}
&& u(x) =  -h \gamma(x,x)\nonumber\\
&& w_1\partial_z \gamma(x,z) + w_2 \partial_x \gamma(x, z) = \int_{x}^{\infty}\gamma(x, y) \hat u(y) \gamma(y,z)dy.
\end{eqnarray}
Similarly, by focusing on the second column of the matrix equation (\ref{dres0}) and defining 
$\hat {\mathfrak G}= {\mathfrak B} {\mathfrak D}^{-1}$ we obtain a second Riccati equation 
\begin{equation}
\hat {\mathfrak G} {\mathrm Y}-{\mathrm X} \hat {\mathfrak G} = \hat u - \hat {\mathfrak G}u \hat {\mathfrak G}. \label{ric2}
\end{equation}
$\hat {\mathfrak G}$ is also an integral operator with kernel $\hat \gamma(x,z)$, 
hence the integral representation of (\ref{ric2}) leads to 
\begin{eqnarray}
&& \hat u(x) = h\hat \gamma(x,x)\nonumber\\
&& w_1\partial_x \hat \gamma(x,z) + w_2 \partial_z \hat \gamma(x, z) = \int_{x}^{\infty}\hat \gamma(x, y) u(y) \hat \gamma(y,z)dy.
\end{eqnarray}
The $t$-evolution of the dressing transformation (\ref{generalDarboux}) naturally provides $t$-Riccati equations for 
${\mathfrak G},\ \hat {\mathfrak G}$, (see also \cite{DFS2}, see also \cite{BDMS, BDMS2} for relevant discussion on non-local non-linear time evolution equations).

With this we conclude our derivation of the non-commutative Riccati equation  associated to the general matrix Darboux 
transform expressed as an ${1\over \lambda}$ formal series expansion as well as an integral operator.

\subsection*{Acknowledgments}
\noindent We are indebted to P. Adamopoulou, V. Caudrelier and G. Papamikos for illuminating discussions. A.D. is grateful to 
S.J.A. Malham and I. Stylianidis for stimulating conversations and ongoing collaboration.
A.D. also wishes to thank LPTM, University of Cergy-Pontoise, where part of this work was completed, and J. Avan for kind hospitality 
and useful comments and suggestions. I.F. is supported by EPSRC funding via a DTA scholarship. S.S. is supported by Heriot-Watt 
funding via a J. Watt scholarship.

\end{document}